\documentclass[conference]{IEEEtran}
\usepackage{color} 
\usepackage{fancyhdr}
\usepackage{cite}

\usepackage{graphicx}
\DeclareGraphicsExtensions{.eps}
\usepackage{setspace}
\providecommand{\PSforPDF}[1]{#1}
\usepackage[cmex10]{amsmath}
\usepackage{amsfonts,amssymb,amsthm}
\usepackage{times}
\PSforPDF{\usepackage{psfrag}}
\usepackage{algorithmic}
\usepackage{algorithm}
\usepackage{array}
\usepackage{balance}
\usepackage{multirow}
\usepackage{amssymb}
\usepackage{mdwmath}
\usepackage{url}
\usepackage{mdwtab}
\usepackage{eqparbox}
\usepackage{nicefrac}
\usepackage{soul,xcolor}
\setstcolor{magenta}
\usepackage{exscale,relsize}
\ifCLASSINFOpdf
\else
\fi
\begin{document}

\title{Energy-Efficient Load-Adaptive Massive MIMO }
%
%\author{\IEEEauthorblockN{M M Aftab Hossain}
%\IEEEauthorblockA{School of Electrical Eng.\\
%Aalto University, Finland\\
%Email: mm.hossain@aalto.fi}
%\and
%\IEEEauthorblockN{Riku J\"antti}
%\IEEEauthorblockA{School of Electrical Eng.\\
%Aalto University, Finland\\
%Email:riku.jantti@aalto.fi}
%\and
%\IEEEauthorblockN{Cicek Cavdar}
%Wireless@KTH, \\
%KTH RIT, Sweden\\
%Email:cavdar@kth.se}
%\and
 %\IEEEauthorblockN{Emil Bj ¨ornson}
%Dept. of Electrical Eng. (ISY)\\ 
%Link ¨oping University, Sweden\\
%Email:emil.bjornson@liu.se

%emil.bjornson@liu.se
%\ vspace * . {- 55cm }}

\author{
\IEEEauthorblockN { M M Aftab Hossain \IEEEauthorrefmark {1},  Cicek Cavdar \IEEEauthorrefmark {2},  Emil Bj\"ornson \IEEEauthorrefmark {3}, and Riku J\"antti \IEEEauthorrefmark {1}} 
\IEEEauthorblockA{\IEEEauthorrefmark{1}School of Electrical Engineering, Aalto University, Finland
\\Email:\{mm.hossain, riku.jantti\}@aalto.fi} 
\IEEEauthorblockA{\IEEEauthorrefmark{2}Wireless@KTH, KTH Royal Institute of Technology, Sweden
 \\Email: \ { cavdar@kth.se }}
\IEEEauthorblockA{\IEEEauthorrefmark{3} Dept. of Electrical Eng. (ISY), Link\"oping University, Sweden
 \\Email: \ { emil.bjornson@liu.se}} }
\maketitle

\begin{abstract}                                       
Massive MIMO is a promising  technique to  meet the exponential growth of global mobile data traffic  demand. However, contrary to the current systems, energy consumption of  next generation networks is required to be load adaptive as the network load varies significantly throughout the day. In this paper, we propose a load adaptive massive MIMO system  that varies the number of antennas following the daily load profile (DLP) in order to maximize the downlink energy efficiency (EE). A multi-cell system is considered where each base station (BS) is equipped with a large number of antennas to serve many single antenna users.  In order to incorporate  DLP, each BS is modeled as an $M/G/m/m$ state dependent queue under the assumption that the network is  dimensioned  to serve a maximum number of users at the peak load. For a given number of users in a cell, the optimum number of active antennas maximizing EE is derived. The EE maximization problem is formulated in a game theoretic framework where the number of antennas to be used by a BS is determined through best response iteration.  This load adaptive system  achieves overall $19$\% higher EE compared to a baseline system where the BSs always run with the fixed number of antennas that is most energy efficient at peak load and that can be switched-off when there is no traffic.
\end{abstract}


\section{Introduction}
The goal of $5$G cellular networks is to provide thousand fold capacity increase while keeping the same cost as today. As the network load varies significantly throughout the day,   it is very important that $5$G networks are capable of adapting their power consumption with this temporal variation of load. Massive MIMO is expected to be a leading candidate technology that can cater very high capacity. In massive MIMO systems, each BS uses hundreds  of antennas to simultaneously serve tens of user equipments (UEs) on the same time-frequency resource \cite{Larsson}.  This study aims to give an insight to the energy efficient design of multi-cell massive MIMO system taking into account the dynamic efficiency of a power amplifier (PA)  and adaptive activation of antennas following the DLP. 

 Recently, both  massive MIMO and energy efficiency of wireless systems  have garnered significant attention \cite{Da,Yang, Emilc, Emilj, Ngoc}.  In \cite{Yang} and \cite{Emilc}, the role of circuit power in EE of massive MIMO has been emphasized. Specifically, in \cite{Emilc}, it  has been shown that without accounting for circuit power consumption, an infinite EE can be achieved as the number of antennas, $M\to\infty$, which is misleading. In \cite{Emilc} and \cite{Emilj}, the authors show that the EE is a quasi-concave function of  the three main design parameters; namely, number of BS antennas, number of users and transmit power. They also show that the energy optimal strategy requires  increasing the transmission power with the number of antennas if the circuit power consumption is taken into account.  In \cite{Ngoc}, an adaptive antenna selection scheme has been proposed where both the number of active RF (radio frequency) chains and the antenna indices are selected depending on the channel condition. However, none of these studies has provided any mechanism to cope with the daily load variation and maintain high EE throughout the day in a multi-cell scenario. 

In this work,  we find the optimum number of BS antennas, $M$, maximizing the EE of a multi-cell massive MIMO system for any given number of users, $K$.  Note that we measure the EE in bit/Joule, i.e., ratio between the average achievable data rate and the total average power consumption~\cite{Emilc, Chen}.  As we want to maximize EE throughout $24$ hour operation of the network, we optimize $M$ following the DLP. In order to map the user distribution to the DLP, we model each BS as a state dependent $M/G/m/m$ queue \cite{Cruz, Cheah} and utilize the DLP as suggested in~\cite{EarthModel}. The $M/G/m/m$ queue dictates that for exponential arrival and general distribution of service time, maximum $m$ number of users can be served simultaneously (number of servers \!=\! $m$, waiting place \!=\! 0). The state dependency arises from the fact that the user rate depends on the number of users the BS serves simultaneously.  We assume the  maximum number of users that a  BS is allowed to serve is $m \! \!= \!\!K_{max}$ and the network is dimensioned in a way that this  $K_{max}$ corresponds to the peak load of the DLP.  We  find  $K_{max}$, corresponding optimum number of antennas, $M_{max}$ and optimum average power per antenna, $p$ that maximize EE when serving the peak load, with the assumption of fixed average transmission power per antenna and considering the realistic efficiency characteristics of non-ideal power amplifier (PA).  Note that under this assumption the network becomes most energy efficient when serving maximum load. As the number of antennas that maximize the EE of a cell depends on the number of antennas used by the interfering cells, we propose a distributed algorithm where the number of antennas is determined through a best response iteration. As the formulated EE maximization problem is not convex with $M$, we resort to a game theoretic approach to achieve the convergence of the proposed algorithm to a Nash equilibrium.

In  the numerical section, we illustrate the potential to increase EE by adapting the number of antennas in a multi-cell massive MIMO system. We observe that the gain in EE is even around 300\% at very low load when compared with a baseline system which does not adapt the number of antennas to the DLP, i.e., each BS uses $M_{max}$ antennas if there is at least one user and turns off all the antennas otherwise.  However, this gain  keeps decreasing with the increase in load. We see that over $24$ hour operation, the overall gain is 19\%, when the day is divided into $24$ intervals, i.e., hourly average load is used as input. 
 
The rest of the paper is organized as follows: in  Section~\ref{sec:SystemModel}, we present the system model and in Section \ref{sec:ProblemFormulation}, we formulate the EE maximization problem. The optimization algorithm based on best response iteration is presented and discussed in Section \ref{sec:ESG}. In Section \ref{sec:NumericalResults}, we illustrate the findings of the numerical analysis. We conclude the paper in Section~\ref{sec:Conclusion}.

\section{System model}
\label{sec:SystemModel}

Let us consider the  downlink of a multi-cell massive MIMO system consisting of cells with indices in the set $\mathcal{C}= \{1, 2, ..., C\}$ and each having its own BS. In the following the terms cell and BS are used interchangeably. The BS $c \in \mathcal{C}$  uses $M_c$ antennas to  serve $K_c$  single antenna UEs. Each antenna of the BS has its own power amplifier. We consider Rayleigh fading channels to the UEs and the spacing between adjacent antennas at the BS is such that the   channel components  between the BS antennas and the single-antenna UEs are uncorrelated. Under the assumption of this independent fading and considering the fact that power gets averaged over many subcarriers,  each antenna uses the same average power. Let us denote  this average transmit power per antenna by $p$, hence, the total transmit power of cell $c$ is  $P_c = p M_c$. The number of active antennas of any other cell $d \ne c$ is $M_d$ and the corresponding transmit power is $P_d=p M_d$. Note that the average transmit power of a BS is not fixed as it varies with the number of active antennas.  The large-scale fading, i.e., the average channel attenuation due to path-loss, scattering, and shadowing from the BS to the UEs is assumed to be the same for all the antennas, as the distance between any UE and the BS is  much larger than the distance among the antennas. 

Let us assume that the BSs and UEs employ a time-division duplex (TDD) protocol and are perfectly synchronized. We also assume that the BS obtains perfect CSI from the uplink pilots, which is a reasonable assumption for low-mobility scenarios. Each BS employs zero forcing precoding so that the intracell interference is canceled out and the power allocation is adapted to make sure that all the users achieve the same average data rate.    Let us denote this rate for the users in cell $c$ by $R_c$. Note that $R_c$ is a function of $K_c$, $M_c$ and $M_d$, $d \in \mathcal{C}$, $d \ne c$. The average data rate achieved by each user in cell $c$ under the above assumptions can be  given by \cite{ Emilmo}  
\begin{equation}
R_c\!=\! B\Big(1\!-\!\frac{K_{max}}{T_c}\Big) \log_{2}\! \Bigg(\!1\!+\!\frac{p \frac{M_c}{K_c}  (M_c-K_c)}{\Lambda_{cc} \sigma^2+ \sum_{d \neq c} \Lambda_{cd} p M_d } \!\Bigg)
\label{eq:Rc}
\end{equation} 
where $T_c$ is the length of the channel coherence interval (in symbols),  $K_{max}$ is the maximum number of users  assumed to be the same for all cells, $\Big(1-\frac{K_{max}}{T_c}\Big)$ accounts for the necessary overhead for channel estimation,  $B$ is the bandwidth,  $\Lambda_{cc} = \mathbb{E}\{\frac{1}{v_{serving}}\}$, where  the random variable  $v_{serving}$ is the channel variance from the serving BS and  $ \sum_{d \neq c} \Lambda_{cd} p M_d$ is the average inter-cell interference power from cell $d$ to $c$  normalized by $\Lambda_{cc}$. Note that $R_c$ is achieved by averaging over the locations of the users of cell $c$. As a result, $R_c$ is a tractable lower bound on the average capacity of a cell.

\section{Problem formulation}
\label{sec:ProblemFormulation}
In this study we want to maximize the EE of the multi-cell massive MIMO system defined in Section \ref{sec:SystemModel}.  The EE is defined as the number of bits transferred per Joule of energy and hence can be computed as the ratio of average sum rate (in bit/second) and the average total power consumption (in Joule/s)~\cite{Emilj}. Power consumed in a BS depends on the number of active antennas and number of users served simultaneously.  If  $P_c^{tot}( K_c, M_c)$ denotes the total energy consumed by a BS when serving $K_c$ number of users simultaneously using $M_c$ antennas and  $R_c(K_c, M_c,  \{M_d\}_{d \neq c})$ denotes the resulting average data rate per user,  the corresponding EE will be
\begin{eqnarray}
\label{EE_c}
\mathrm{EE} &=&  \frac{{\text{Average sum rate}}}{\text{Power consumption}} = \frac {K_c R_c(K_c, M_c,  \{M_d\}_{d \neq c})} {P_c^{tot}(K_c, M_c)}. \nonumber     
\end{eqnarray} 
The EE maximization problem for cell $c$ for a particular load can be expressed in the following way:
\begin{subequations} 
\label{eq:pproblem1}
\renewcommand\theequation{\theparentequation\roman{equation}}
\begin{align} 
   {\mathop {{\text{maximize}}}_{M_c} } &   
    \frac {K_c R_c(K_c,  M_c, \{M_d\}_{d \neq c}) } {P_c^{tot}(K_c, M_c)}\label{eq:P11}\\ 
 {{\text{subject to}}} &\;\; { K_c+1 \le M_c}  \label{eq:P12}  
\end{align}
\end{subequations}
where $M_d$ is the number of the antennas used by any other cell $d$, $d \ne c$, and the constraint comes from the requirement of zero forcing precoding.
However, the network loads vary throughout the day. In order to capture the daily load variation and maximize EE throughout the day, we model each BS as an $M/G/m/m$ state-dependent queue.  Let us consider that  during time interval $h$, the steadystate probability of the BS $c$ serving $n$ number of users, i.e., $Pr[K_c=n]$  is denoted by $\pi_c(h,n)$. Our objective is to maximize EE by adapting the number of active antennas for any user states taking the received interference into account. The main problem formulation for BS $c$ can be rewritten as
\begin{subequations} 
\label{eq:problem}
\renewcommand\theequation{\theparentequation\roman{equation}}
\begin{align} 
   {\mathop {{\text{maximize}}}_{\mathbb{M}_c}} &\;\;
   {\sum_{h=1}^{H}}
     \sum_{n=1}^m \!\!\pi_c(h, n) \frac {n R_c(n, M_c^{(h)}, \{M_d^{(h)}\}_{d \neq c}) } {P_c^{tot}(n, M_c^{(h)})}  \label{eq:Q11}\\ 
   {{\text{subject to}}} &\;\; { n+1 \le \mathcal{M}_c^{(h)}(n)\le M_{max}} \label{eq:Q12}  
\end{align}
\end{subequations}
where $R_c(n, M_c^{(h)}, \{M_d^{(h)}\}_{d \neq c}) $ is the average rate per user when there are $n$ number of users in the cell at time interval $h$. ${\mathbb{M}_c} = [\mathcal{M}_c^{(1)} \mathcal{M}_c^{(2)} . . . \mathcal{M}_c^{(H)} ]$ where $\mathcal{M}_c^{(h)}$ is the vector that gives the number of antennas that maximize EE at different user states in cell $c$ during the time interval $h$. The constraints come from the requirement of zero forcing. Note that EE optimization could be divided into any arbitrary number of time intervals, $H$. In this work, we set $H=24$ in the simulations in order to optimize the system over $24$ hour operation where  we use the hourly average network load as input. As the solution for  the problem \eqref{eq:problem}  depends on the actions taken at other BSs, we formulate the joint optimization  under a game theory framework. 

\subsection{Traffic model} 
 \label{Sec:TrafficModel}
The maximal number of users that can be served simultaneously is derived based on the queuing model and is denoted by  $m\! = \!K_{max}$. Users achieve different data rates as the number of users served by the BS changes. The network is assumed to be  dimensioned in a way that the data carried by the cell while serving $K_{max}$ corresponds to the peak load of the DLP. In a queuing system with no buffer space, the blocking probability is equal to the probability of having the system 100\% loaded, i.e., probability of having $K_{max}$ number of users. This can be explained by the PASTA (Poisson Arrivals See Time Averages) property which holds when the arrivals are following the Poisson process. In this work, we allow at most $2$\% blocking at peak load, i.e., the probability of serving  the maximum allowed number of users, $K_{max}$ simultaneously is $0.02$.  In order to capture the daily traffic variation, we consider the  DLP proposed for data traffic in Europe  \cite{EarthModel}.   The steady state  probabilities for the random number of users $K$ in the BS $c$ modeled as  $M/G/m/m$ state-dependent queue, $\pi_c(n)\equiv Pr[K_c\!=\!n]$, are as follows: \cite{ Cruz}
\begin{equation}
\pi_c(n)\!=\!\!\left[\frac{\left[\lambda\frac{s}{R_c(1)}\right]^n}{n! f(n)f(n-1)...f(2)f(1)} \right]\pi_c(0), n \!= \!1,2,...m, \nonumber
\end{equation}
where $\pi_c(0)$ is the probability that there is no user in cell $c$ and is given by
\begin{equation}
\pi_c^{-1}(0)=1+ \sum_{i=1}^m\left(\frac{\left[\lambda\frac{s}{R_c(1)}\right]^i}{i! f(i)f(i-1)... f(2) f(1)}\right)  \nonumber
\end{equation}
where $\frac{s}{R_c(1)}$ is the expected service time when BS $c$ serves  a single user, $f(n)=R_c(n)/R_c(1)$ where $R_c(n)$
 is the  data rate per user while serving $n$ number of users, $\lambda$ is the arrival rate, and $s$ is the total data traffic contribution by a single user. Note that we use (\ref{eq:Rc}) to find the data rates at different user states.
\begin{figure}[H] 
\centering
        \includegraphics[ height=1.8in]{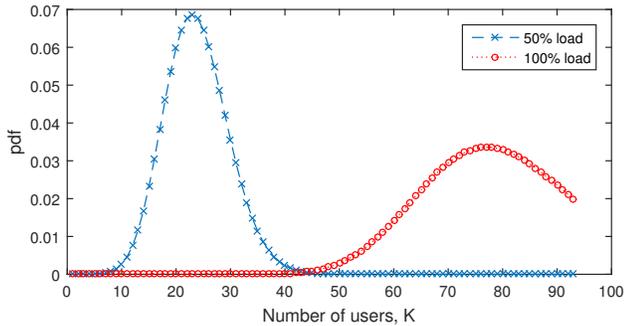}
\caption{User distribution while serving $50$\% and $100$\% cell load with the parameters given in Section \ref{sec:NumericalResults}.}
	\label{fig:Mgcc}
\end{figure}
 In order to find the steady state probability distribution throughout the day, first we set the values for $\lambda$ and $s$. As we allow 2\% blocking rate  while serving $100$\% load, we find the maximum $\lambda$, i.e., $\lambda_{max}$ that results $\pi(K_{max})=0.02$ for a fixed $s$. Assuming that $s$ remains constant, we derive the hourly average number of users following the DLP using $\lambda_{max}$. For example, from the DLP, if the average load at any time interval $h$ is  $x$\%, the corresponding average number of users $\lambda_h= \frac{x}{100}\cdot \lambda_{max}$. This  $\lambda_h$ has been used as the input to the $M/G/m/m$ queue to find the steady state probability distribution  of the users during time interval $h$. 
Fig. \ref{fig:Mgcc} gives two example plots of the user distribution for 50\% and 100\% loads with the parameters provided in Section \ref{sec:NumericalResults}. Note that for 100\% load, the probability of serving $K_{max} =93$ users is $0.02$.

\subsection{Power consumption model}                                  
The total power consumed in a BS is given by 
\begin{equation} 
P_c^{tot}(K_c, M_c)= M_c P_{PA} (p )+P_\mathrm{BB} (K_c, M_c)+P_\mathrm{Oth} \nonumber
\end{equation}
where $P_{PA}(p)$ gives the power consumption of a PA when the average output power is $p$, $P_\mathrm{BB} (K_c, M_c)$ is the base band signal processing power when the BS serves $K_c$ number of users simultaneously with $M_c$ number of antennas. $P_\mathrm{Oth}$ includes the load-independent power for site cooling, control signal, DC-DC conversion loss, etc. For baseband and fixed power consumption we use the model proposed in \cite{Emilc}. The total circuit power is given by 		
\begin{equation} 
\label{eq:Pcp}
P_\mathrm{CP}=P_\mathrm{TC}+P_\mathrm{CE}+P_\mathrm{C/D}+P_\mathrm{LP}. 	\nonumber
\end{equation}				
The power consumed in the transceiver is given by $P_\mathrm{TC}=M P_\mathrm{BS}+P_\mathrm{SYN}$  where $P_\mathrm{BS}$ is the power required to run the circuit components,  e.g., converters, mixers and filters attached to each antenna at the BS and $P_\mathrm{SYN}$ is the power consumed by the local oscillator. $P_\mathrm{CE}$ is the power required for channel estimation process. $P_\mathrm{C/D}$ is the total power required for channel coding, $P_\mathrm{COD}$  and channel decoding, $P_\mathrm{DEC}$.   $P_\mathrm{LP}$ is the power consumed for linear processing.
According to \cite{Emilc}, the total baseband power can be expressed as 
\begin{equation} 
\label{eq:Pbb}
P_\mathrm{BB} (M_c, K_c) = \underbrace {\mathcal{A} K_c R_c+\sum_{i=0}^3 C_{0,i} K_c^i}_{C_0^{BB}}+ M_c \underbrace{\sum_{i=0}^2 C_{1,i} K_c^i}_{C_1^{BB}}  \nonumber
 \end{equation}
where  $C_{0,0}=P_\mathrm{SYN},  C_{0,1}=0,  C_{0,2}=0, C_{0,3}=\frac{B}{3 T_c L_\mathrm{BS}}, C_{1,0}= P_\mathrm{BS}, C_{1,1}=\frac{B}{L_\mathrm{BS}}(2+\frac{1}{T_c}), C_{1,2}=\frac{3B}{L_\mathrm{BS}}, \mathcal{A}=P_\mathrm{COD}+P_\mathrm{DEC}$, $R_c$ is the rate achieved by a user on average as given in equation (\ref{eq:Rc}) and $B$ is the  bandwidth. Note  that the power consumption  $C_0^{BB} $ is independent of the number of antennas used. 
 
 For traditional PAs, the total input power needed for mean output  transmission power $p$ can be approximated as~\cite{Persson,Hossain1} 
\begin{equation}\label{eq:T-PA}
P_{PA}(p)\approx \frac{1}{\eta} \sqrt{p\cdot P_{max,PA}} \nonumber
\end{equation}
where $\eta$ denotes the maximum PA efficiency when transmitting the maximum output power $P_{max,PA}$. Note  that the maximum mean transmit power, $p_{max}$ must be around $8$ dB less than $P_{max,PA}$ due to the high peak to average power ratio (PAPR) of recent technologies, e.g., OFDM.  The total power consumption at BS $c$ can be written as 
\begin{equation}
\label{eq:ETPAeff}
P_{total} \approx C_0+C_1 M_c 
\end{equation}
 where $C_0=C_0^{BB}+P_{Oth}$ and $C_1=C_1^{BB}+P_{PA}(p)$.
	%%%%%%%%
	
\section{Best response iteration and algorithm} \label{sec:ESG}
The objective function in ~\eqref{eq:problem} involves summation over all the user states at different hours. Let us rewrite the problem (\ref{eq:problem}) using  \eqref{eq:Rc} and \eqref{eq:ETPAeff} as
\begin{subequations} 
\label{eq:ProblemLong}
\renewcommand\theequation{\theparentequation\roman{equation}}
\begin{align} 
   {\mathop {{\text{ maximize}}}_{\mathbb{M}_c}} &\;\;
   {\sum_{h=1}^{H}}
     \sum_{n=1}^{m} \!\!\pi_c(h, n) \frac{n \beta \log{({1-n M_c \gamma_{c,1}+\gamma_{c,1}M_c^2})}}{C_0+C_1 M_c} \label{eq:P21}\\
    {{\text{subject to}}} &\;\; { n+1<M_c(n)< M_{max}}   \label{eq:P22}  
\end{align}
\end{subequations}
where  $\gamma_{c,1} =\frac{\frac{1}{n}p}{(\Lambda_{cc} \sigma^2+ \sum_{d \neq c} \Lambda_{cd} p M_d )}$ is the achieved signal to interference plus noise ratio (SINR) by cell $c$ for using a single antenna and $\beta = (1-\frac{K_{max}}{T_c}) \frac{B}{\text{ln}2}$. 
Note that we drop $h$ from notation $\mathcal{M}_c^{(h)}$ henceforth as the optimization for different time interval $h$ can be carried out separately. The objective function  when the BS is serving a particular number  of users, $n$  can be broadly written as 
\begin{equation}
E_c  = \frac  {n \beta\log{({1-n M_c \gamma_{c,1}+\gamma_{c,1}M_c^2})}} {C_0+C_1 M_c}. 
\label{eq:Ebfinal}
\end{equation}
As the transmit power of a BS depends on the number of active antennas, the number of active antennas for different cells are coupled due to inter-cell interference. As a result, the objective function in \eqref{eq:ProblemLong} for cell $c$ is dependent on the number of antennas used by other cells and there is no closed form expression for $M_c$. Because of that we resort to an algorithm based on best response iteration in a game theoretic framework. 
In this framework, each BS  iteratively  finds the most energy efficient number of active antennas taking into account the interference from the surrounding BSs. In order to formulate \eqref{eq:ProblemLong}, in a game theoretic framework,   we start by defining the set of feasible number of antennas at different user state  in cell $c$,  $\mathcal{U}_c= \{1, 2,  ..., m\}$. The set $\mathcal{S}_c$ of feasible number of antennas for the cell  $c$ is a function of the  number of antennas at the interfering BSs, ${\mathbf{M}}_{-c}$.  
\begin{equation}
\label{eq:strategy}
\mathcal{S}_c(\mathbf{M}_{-c})\!=\!\left\{\mathcal{M}_c(n) \!: \!n\!+\!1\!\! \leq \!  \mathcal{M}_c(n) \!\leq\! M_{max}, \!\! \forall n\in\mathcal{U}_c\right\}. 
\end{equation}
Next, we  define the {\em EE maximization game}, $\mathcal{G}(\mathcal{K},\mathcal{S},\mathcal{E})$ where the players are the BSs,  $S=S_1 \times S_2 \times \cdots, S_C$ is the strategy space, i.e., space of number of active antennas, and ${\mathcal{E}}=E_c(\mathbf{M}_{c},\mathbf{M}_{-c}), c\in\mathcal{C}$ is the utility of the players, i.e., EE  of the cells. 
 The {\em best response} is the strategy (or strategies) that produces the most favorable outcome for a player given other players' strategies. The use of best response strategy gives rise to a dynamic system of the form
\begin{equation}
\label{eq:maximize}
\mathcal{M}_c={\arg\max}_{\mathcal{M}_c\in\mathcal{S}_c(\mathbf{M}_{-c})} 
{E}_{c}(\mathcal{M}_c,\mathbf{M}_{-c}), \forall c. 
\end{equation}

\begin{algorithm}
\label{algo}
\caption{Best response iteration}
\begin{algorithmic}
\STATE ${\mathcal{M}}_{c} \leftarrow M_{max} \cdot \mathbf{1}, \forall c\in\cal{C}$. 
\STATE $\textnormal{maxtol} \leftarrow 1 $
\WHILE{$\textnormal{maxtol} \neq 0$} 
\FORALL{$c\in\cal{C}$}
\STATE $i\leftarrow c$ 
\STATE ${\mathbf{M}}_{-c}\leftarrow \sum_n{\mathcal{M}_d(n)\pi_n}, \forall n\in {\cal{U}}_{d},\forall d: d \neq c$ 
\STATE Define strategy space $\mathcal{S}_c$ based on~\eqref{eq:strategy}
\STATE $\mathcal{M'}_c \leftarrow {\arg\max}_{\mathcal{M}_c\in\mathcal{U}_c(\mathbf{M}_{-c})}
{E}_{c}(\mathbf{M}_{-c})$
\STATE $\mathrm{tol}_c\leftarrow |{\mathcal{M'}}_{c} - {\mathcal{M}}_{c}|$
\STATE $\mathcal{M}_c \leftarrow \mathcal{M'}_c$
\ENDFOR
\STATE $\textnormal{maxtol} \leftarrow {\text{max}}_{c}(\mathrm{tol}_c) $
\ENDWHILE
\end{algorithmic}
\end{algorithm} 
 We present the best response optimization algorithm in the form of pseudocode, see Algorithm $1$.  Initially, the number of antennas of all the cells are set  to the maximum, $M_{max}$. We start the best response iterations from the $c$-th cell. The interference level received by any user in the cell  $c$, $I_c({\mathbf{M}}_{-c})$, is a function of the number of antennas used in the other cells, $\mathbf{M}_{-c}$. In order to compute the interference level  $I_c(\mathbf{M}_{-c})$, it is assumed that the $d$-th interfering BS, $d\neq c$, transmits with the number of antennas found from the weighted mean of the number of antennas for its different user states i.e., $\sum_{n=1}^m{\mathcal{M}_d(n) \pi_d(n)},  \forall d\!:\!d\neq c$. Once the interference caused to the  $c$-th cell is calculated, we identify the strategy space for the $c$-th cell based on equation~\eqref{eq:strategy}. Finally, we find the vector of antennas that maximizes EE at different user states, $\mathcal{M'}_c \leftarrow {\arg\max}_{\mathcal{M}_c\in\mathcal{S}_c(\mathbf{M}_{-c})} E_c(\mathbf{M}_{-c})$ where '$\leftarrow$' indicates the direction of value assignment. We have to iterate over all the cells and  optimize the antenna vector for each cell. The iterations are carried out until the antenna vector for each cell converges, i.e., there is one iteration where none of the antenna numbers changes. 

The number of antennas for a user state $n$,  $M_c(n),  n\in {\mathcal{U}}_c$, is independent of the antennas used  for the other user states in the same cell. Therefore, in order to carry out the optimization step, $\mathcal{M'}_c \leftarrow {\arg\max}_{\mathcal{M}_c\in\mathcal{S}_c(\mathbf{M}_{-c})} {E}_{c}(\mathbf{M}_{-c})$, it is sufficient to solve the optimization problem~\eqref{eq:ProblemLong} separately for each user state. For the given number of antennas of the interfering BSs, the interference is known. For known interference the EE problem for any user state is a quasi-concave function of $M_c$ as it is a ratio of a concave and an affine function of $M_c$ \cite{Ratio}. As a result, it suffices to compute the value of the objective function at the stationary point and at the end points of the interval to identify the optimal $M_c$. 
 
In order to prove the convergence  of the best response strategy  for dynamic adaptation of the number of antennas to a Nash equilibrium, it can be shown that  the objective function has {\em increasing differences} in $(\mathcal{M}_c,\mathbf{M}_{-c})$ which enables us to formulate the problem as a {\em super-modular game} \cite{Topkis}. Therefore  the best response converges to a {\em Nash equilibrium} which is unique \cite{Altman}.

%%%%%%%%%%%%%%%%%%%%%%%%%%%%%%%`%%%%%%%%
\section{Numerical analysis}
\label{sec:NumericalResults}
We consider the downlink of a cellular network with 19 regular  hexagonal cells where the wrap around technique is applied in order to get rid of the boundary effect. The maximum cell radius is $500$ m if not otherwise specified. The users that reside in the inner $35$ m from the cell center are not considered in this analysis  and a uniform user distribution has been assumed outside that range. Note that we consider $15000$ test points in each cell in order to calculate the average channel variance from the serving BS, $\Lambda_{cc}$ and the average inter-cell interference power, $\Lambda_{cd}$. We also consider uniform load for the cells.  The parameters for the simulation are given in  Table \ref{RP}. Some of them are taken from~\cite{Emilj}.

\begin{table}[!ht]
\caption{Simulation parameters}
\label{RP} \centering
\begin{tabular}{|l|l|} \hline
\multicolumn {2} {|c|}{\textbf {Reference parameters }} \\
 \hline
  \textit{Parameter} & \textit{Value} \\
 \hline
 Number of cells & $19$ \\
 \hline
Grid size inside each cell & $15000$ points \\
 \hline
Cell radius: $d_{max}$ & $500$ m \\
\hline
Minimum distance: $d_{min}$ & $35$ m \\
 \hline
Maximum PA efficiency  & $80$\%\\
 \hline
Path loss at distance d & $\frac{10^{-3.53}}{||d||^{3.76}}$\\
 \hline
Local oscillator power: $P_\mathrm{SYN}$  & $2$ W\\
\hline
BS circuit power: $P_\mathrm{BS}$ & $1$ W \\
\hline
Other power: $P_\mathrm{Oth}$  & $18$ W\\
\hline
Power for data coding: $P_\mathrm{COD}$  & $0.1$ W/(Gbit/s) \\
\hline
Power for data decoding: $P_\mathrm{DEC}$  & $0.8$ W/(Gbit/s) \\
\hline
Computational efficiency at BSs:$L_\mathrm{BS}$ & 12.8 Gflops/W \\
\hline
Bandwidth  & $20$ MHz\\
 \hline
Total noise power: $B\sigma^2$  & -$96$ dBm\\
 \hline
Channel coherence interval: $T_c$  & $1800$ symbols \\
\hline
\end{tabular}
\end {table}

For the parameters given in Table \ref{RP}, the optimum value of the transmit power per antenna is found to be  $0.098$ Watt. In order to compare, a reference system has been considered  where  the number of antennas that maximize the EE during 100\% cell load are  kept  active disregarding the number of users that the BS serves and are turned off altogether if the BS serves none.   In the following, first we illustrate how $M_c$ changes with $K_c$ at different cell loads. Then we present the overall performance in terms of rate and EE  which has been generated by taking the weighted average over the performance achieved for all the user states at each particular average cell load. 

\subsubsection{Interplay between number of active users and active antennas in the system}
\begin{figure}[!h] 
\centering
        \includegraphics[ scale=0.65]{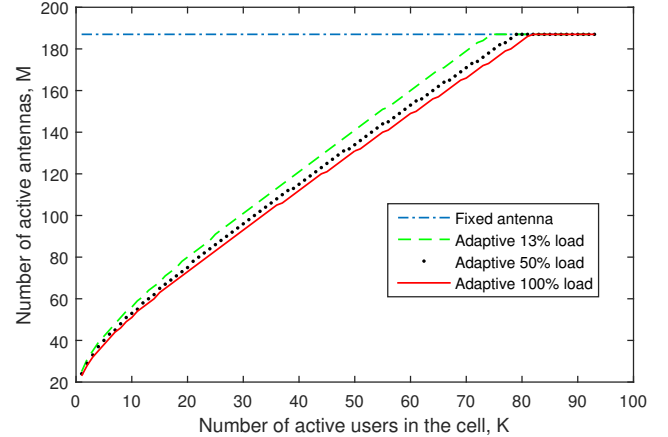}	
\caption{	Number of antennas as a function of the number of users in the cell at different average cell load.}
\label{fig:AntennaStates}
\end{figure}

The probability of getting different number of  users served by the BS simultaneously depends on the average cell load, see Fig. $1$. In Fig.~\ref{fig:AntennaStates}, we show how the number of antennas increases with the number of active users in a cell at different average cell load. Note that for the reference case, all the available antennas are kept active except for the case when there is no user in the system. However, when the EE is optimized over changing number of users, the number of active antennas adapts to the user profile. The ratio between the number of antennas and the number of users is higher than three when there are few users in the system and ends up slightly higher than two at high load. Overall, the relation between the number of antennas and the number of users is quite linear. 

\subsection{Average rate per user at different cell load}
\begin{figure}[!h] 
\centering
        \includegraphics[ scale=0.65]{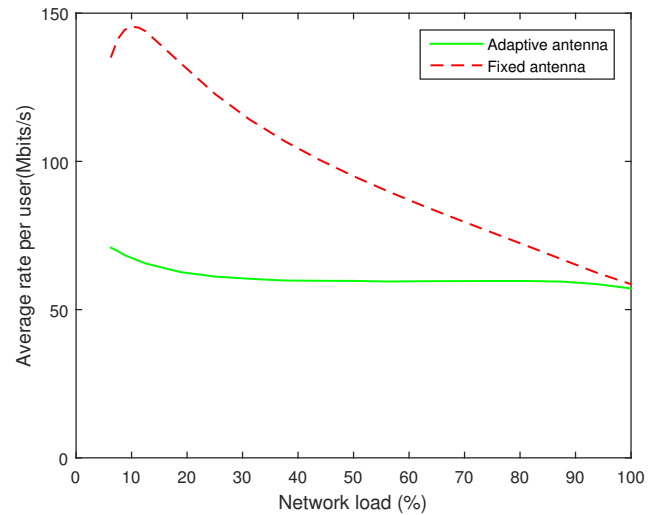}	
\caption{Average rates in a cell at different cell loads.}
\label{fig:RealRates}
\end{figure}

In Fig.~\ref{fig:RealRates}, we show the average rate achieved by a user at different network load. It is observed that when the network load is low, the user rates decrease considerabely compared to the reference case.  However, the gap reduces with the increase of network load.

\subsection{EE gain at different cell load}
\begin{figure}[!h] 
\centering
        \includegraphics[ scale=0.65]{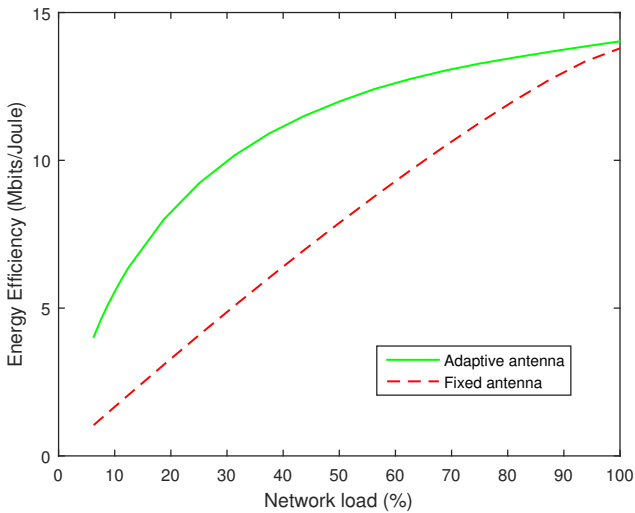}	
\caption{Average EE gain at different cell load.}
\label{fig:EEOverall}
\end{figure}

Fig.~\ref{fig:EEOverall} shows the overall gain in EE as the average load of the network increases.   The largest percentage gain is achieved at very low load and then the gain decreases as the load increases. At peak load, the gain is insignificant as the gain from the probability of having small number of users that allow EE improvement by reducing antennas is very low,  see Fig. $1$.

\balance
\subsection {EE and user rate tradeoff}
\begin{figure}[!h] 
\centering
        \includegraphics[ scale=0.65]{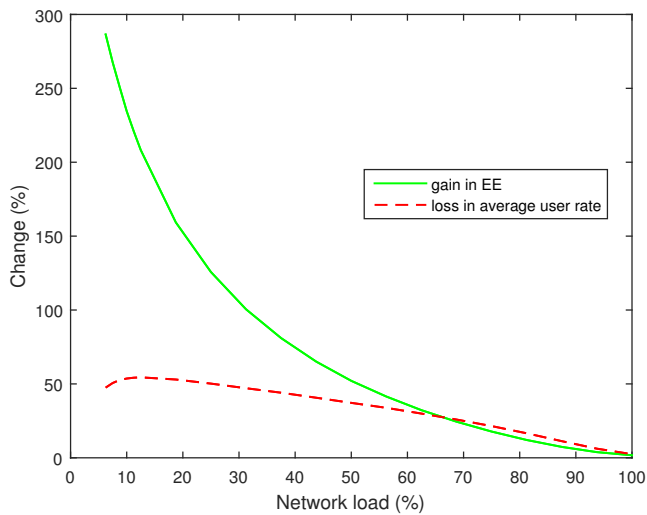}	
\caption{Gain in EE at the cost of rate performance.}
\label{fig:TradeOff}
\end{figure}

Fig.~\ref{fig:TradeOff} shows the tradeoff between the EE and the average user rate at different load. At very low load the EE has been increased around 300\% at the cost of around 50\% reduction of average user date rate. However, with the increase of load in the system, both the gain in EE and loss of user rate get reduced. 

\section{Conclusion}
\label{sec:Conclusion} 
In this work we investigate how to dynamically adapt  massive MIMO systems  to the user loads for higher EE. The temporal variation of load has been captured by modeling each BS using massive MIMO  with an $M/G/m/m$ queue and mapping the user distribution to a DLP. We developed a game theory based distributive algorithm that yields significant gains in EE at the cost of reduction of average user data rate at low user load. However, the high rate degradation while increasing the EE  comes from the fact that we consider a very tight reference case. In our reference case, the system considers the complete shutdown of all the antennas when the BS is not serving any user. This reduces the interference significantly resulting high data rates for users which in turn allows the BS to reduce its activity time further.  For transparency, the algorithm was developed for a simple rate formula based on perfect CSI, but the same methodology can be applied to other rate formulas as well.

\section*{Acknowledgment}
This work was  supported by EIT ICT Labs funded  5GrEEn, EXAM;   Academy of Finland funded  FUN5G, and also by ELLIIT and the CENIIT project.


\begin{thebibliography}{1}

%\bibitem{Cisco} "Cisco Visual Networking Index: Global Mobile Data Traffic Forecast Update, 2013-2018"', Feb 2014
\bibitem{Larsson}
Larsson E. G.,  Tufvesson F., Edfors O., and Marzetta T. L., ``Massive MIMO for Next Generation Wireless Systems," \emph{IEEE Commun. Mag.}, vol. 52, no. 2, pp. 186-195, Feb. 2014.

\bibitem{Da} Feng D. \emph{et al}., ``A survey of energy-efficient wireless communications,"  \emph{IEEE Commun. Surveys \& Tutorials}, vol.15, no.1, pp.167-178, 2013.
	
\bibitem{Yang}	 Yang H. and  Marzetta T. L., ``Total energy efficiency of cellular large scale antenna system multiple access mobile networks," in \emph{Proc. IEEE Online GreenComm}, 2013.
		
\bibitem{Emilc} Bj\"ornson E., Sanguinetti L., Hoydis J., Debbah M., ``Designing Multi-User MIMO for Energy Efficiency: When is Massive MIMO the Answer?,'' in Proc. of \emph{IEEE  WCNC}, Istanbul, Turkey, April 2014.
			
\bibitem{Emilj}  Bj\"ornson E.,  Sanguinetti L.,  Hoydis J.,  Debbah M.,``Optimal Design of Energy-Efficient Multi-User MIMO Systems: Is Massive MIMO the Answer?," \emph{ IEEE Trans.  Wireless Commun.}, vol. 14, no. 6, pp. 3059-3075, June 2015.
		
\bibitem{Ngoc} Le N. P.; Tran L. C.; Safaei F., ``Adaptive antenna selection for energy-efficient MIMO-OFDM wireless systems," in \emph{International Symposium on WPMC}, 2014 , vol., no., pp. 60-64,  Sept. 2014.
	
%\bibitem{Emils} Björnson E., Sanguinetti  L.,  Kountouris M., ``Deploying Dense Networks for Maximal Energy Efficiency: Small Cells Meet Massive MIMO," \emph{IEEE J. Sel. Areas Commun.}, Submitted for publication.
	
\bibitem{Chen}  Chen Y.,  Zhang S.,  Xu S., and  Li G., ``Fundamental trade-offs on green wireless networks'', \emph{IEEE Commun. Mag.}, vol. 49, no. 6, pp. 30-37, 2011.
		
\bibitem{Cruz} Cruz F. R.B, Smith J. M., `` Approximate analysis of M/G/c/c state Dependent Queeing networks,'' \emph{Computers and Operation research}, vol. 34, Issue 8, pp. 2332-2344, 2007.

\bibitem{Cheah} Cheah J. Y., Smith J. M., "Generalized M/G/C/C state dependent queueing models and pedestrian traffic flows," \emph{Queueing Syst.},  Vol. 15, Issue 1-4, pp. 365-386, 1994. 
	
\bibitem{EarthModel} Auer G. \emph{et al.}, ``D2.3: Energy efficiency analysis of the reference systems, areas of improvements and target breakdown," Energy Aware Radio and Network Technologies (EARTH) INFSO-ICT-247733, ver. 2.0, 2012. [Online]. Available: \url{http://www.ict-earth.eu/}

	
\bibitem{Emilmo} Bj\"ornson E. \emph{et al}.,	``Multi-Objective Signal Processing Optimization: The Way to Balance Conflicting Metrics in 5G Systems"  to be published in \emph{IEEE Signal Process. Mag.}, vol. 31, no. 6, pp. 14-23, Nov. 2014.

\bibitem{Hossain1}  Hossain M. M. A. and J\"antti R., ``Impact of efficient power amplifiers in wireless access'', in \emph{IEEE online GreenComm} pp. 36-40, Sept. 2011.
	
\bibitem{Persson} Persson, D.; Eriksson, T.; Larsson, E.G., ``Amplifier-Aware Multiple-Input Multiple-Output Power Allocation," \emph{ IEEE Commun. Letters},  vol. 17, no. 6, pp. 1112-1115, June 2013.

\bibitem{Ratio}  Gotoh J., Hiroshi Konna H., ``Maximization of the Ratio of Two Convex Quadratic Functions over a Polytope", \emph{Computational Optimization and Applications}, Volume 20, Issue 1, pp. 43-60, October 2001.

\bibitem{Topkis}  Topkis D. M., \emph{Supermodularity and Complementarity}.    Princeton, NJ, Princeton Univ. Press, 1998.
%\bibitem{Hossain} Hossain, M.M.A; Koufos, K.; J\"antti, R., ``Minimum-Energy Power and Rate Control for Fair Scheduling in the Cellular Downlink under Flow Level Delay Constraint," \emph{IEEE Trans. Wireless Commun.},   vol. 12, no. 7, pp. 3253-3263, July 2013.

\bibitem{Altman}  Altman E.,  Altman Z., ``S-modular games and power control in  wireless networks," \emph{IEEE Trans. on Automatic Control}, vol. 48, no. 5, pp. 839-842, May 2003.


	
\end{thebibliography}
\end{document}